# Defects of the Crystal Structure and Jahn-Teller distortion in BiMnO$_3$

D.W. Boukhvalov and I. V. Solovyev

*Computational Materials Science Center, National Institute for Materials Science,*

*1-2-1 Sengen, Tsukuba, Ibaraki 305-0047, Japan*

*Using density-functional theory with the on-site Coulomb correction (the LDA+U method), we perform the structural optimization of BiMnO$_3$ by starting from different experimentally reported structures. We confirm that irrespectively on the starting condition, all calculations converge to the same centrosymmetric structure, in agreement with the previous finding. Nevertheless, the structural optimization substantially reduces the Jahn-Teller (JT) distortion in the system. We attribute this fact to the strong competition of local distortions around the Mn- and Bi-sites: while the local Mn-environment experiences the JT instability, the one of the Bi-sites favours the off-centrosymmetric displacements, which involves the same oxygen atoms. The existence of the second mechanism explains the difference between BiMnO$_3$ and more canonical JT manganites, such as LaMnO$_3$. Finally, being motivated by experimental studies, we have investigated the formation of different types of defects and obtained that BiMnO$_3$ (contrary to other considered systems, such as LaMnO$_3$ and BiFeO$_3$) can relatively easily form oxygen impurities at interstitial sites. The impurity oxygen atom tends to form a pair with the host oxygen, which explains the insulating character of the oxygen-excessive BiMnO$_{3+x}$. Moreover, we found that the BiMnO$_{3+x}$ samples experience the "memory effect", where the optimized crystal structure strongly depends on the starting configuration. We suggest that such a memory effect may explain stability of some of the crystal structures of BiMnO$_3$, which have been previously reported experimentally.*

# 1. Introduction

Multiferroics is a relatively new class of materials with unusual properties, which is intensively studied in recent years [1-4]. The coexistence magnetism and ferroelectricity makes possible the application of these compounds for transformation of magnetisation to electric voltage and back. In contrast to the large number of multiferroics, $BiMnO_3$ is a ferromagnet with the Curie temperature of about 100 K and the saturation magnetization 3.92 $\mu_B$/Mn [5]. Earlier structural measurements reported by Atou et. al [6] and dos Santos et. al. [7] revealed two noncentocymmetric structures (in the following, denoted as the A and D structures), crystallizing in the space group the C2. However, heoretical calculations based on the density functional theory (DFT) predict relaxation of both these structures to the single centrosymmetric structure with the space groups C2/c [8]. Furthermore, structural measurements, performed by Belik and co-workers [5], reported two new centrosymetric structures in the low- and high-temperature regime (in the following, denoted as the LT and HT structures). Both structures had the monoclinic C2/c symmetry. Under the pressure, $BiMnO_3$ can be transformed to the orthorhombic Pnma structure, similar to $LaMnO_3$ [9]. Several oxygen-rich $BiMnO_3$ structures had also been reported in the literature [10]. Because of such 'structural multiplicity' caused by the defects, it is currently impossible to grow the single crystal of $BiMnO_3$ [5]. All reported structural data had been collected for X-ray powder diffraction and very sensitive to the conditions of the samples preparation.

The explanation of the experimentally observation ferroelectricity in $BiMnO_3$ remains a challenge to theorists. To this end, the lone pairs model [Spaldin] and improper ferroelectricity caused by the long-range antiferromagnetic interactions in the otherwise centrosymmetric $BiMnO_3$ structure [11, 12] have been proposed. In the present work, we explore the stability of all experimentally detected structural phases and discus the leading mechanisms of structural distortions in the $BiMnO_3$ perovskite. The stabilization of different structures by defects is also discussed.

## 2. Computational method

The electronic structure calculations and structural optimization have been performed by using the DFT-based [13, 14] pseudopotential 'pwscf' code, which is part of the *Quantum ESPRESSO* package [15]. All calculations have been performed for the ferromagnetic configuration and include the correction for the on-site Coulomb repulsion (the so-called LDA+U method) [16, 17]. We have used PBE parameterization [18] for the ultrasoft GGA pseudopotentials [19], the energy cut-offs of 75 and 400 Ry for the plane-wave expansion of the wavefunctions and the charge density, respectively, and a $2 \times 2 \times 2$ Monkhorst–Pack *k*-point grid for the Brillouin sampling [20]. The density of states was calculated on the grid of the $8 \times 8 \times 4$ *k*-point grid, using the smearing of 0.1 eV.

In order to test the choice of the pseudopotentital and technical parameters we have performed the full structural optimization of two similar compounds LaMnO$_3$ [21] and BiFeO$_3$ [22, 23]. For the obtained lattice parameters, the difference from the experimental values was within 1%. The total energy difference between ferromagnetic and antiferromagnetic configurations and the values of the Jahn-Teller distortion for LaMnO$_3$ were close to the previous theoretical studies [24].

Formation energy is calculated by standard formula: $E_{form} = E_{defect} - (E_{pure} \pm E_{impurity})$, where $E_{defect}$ is the total energy of system with defect, $E_{pure}$ is the total energy of system without defects, and $E_{impurity}$ is the total energy of impurity atoms in the bounded state (correspondingly, antiferromagnetic α-Mn, α-Bi, bulk La, and O$_2$ molecule in the triplet state of gaseous phase). The structural optimization was taken into account in all three stages of these calculations.

## 3. Crystal structure and Jahn-Teller distortions in BiMnO$_3$

BiMnO$_3$ has strongly distorted perovskite structure. The values of ratio the longest Mn-O distance in different types of BiMnO$_3$ presented in the Table I. For all structures except high temperature from ref. [5] these values are bigger then corresponding values in LaMnO$_3$ [21]. In order to study stability of different structural configurations of BiMnO$_3$ we have performed optimisation of all structures within GGA and GGA+U with value of Coulomb repulsion 2, 4 and 8eV. Similarly to the previous theoretical work [8] we have found that, irrespectively on the way of the optimization, all earlier proposed A and D non-centrosymmetric crystal structures systematically converge to the same centrosymmetric optimized structure (thereafter denoted as OPT), which is close to the HT monoclinic structure (although, there are some differences, which will be discussed below). The LT monoclinic structure, after the relaxation, also converges to the OPT structure. In all cases, the structural optimization systematic decreases the Jahn-Teller (JT) distortion (Tables I-III).

Actually, the ability of GGA to describe correctly the JT distortion in transition-metal oxides was one of the controversial and hotly debated issues in the middle of 1990$^{th}$ [17, 25]. Although the JT effects is primarily caused by the electron-lattice interactions in the system [26], which formally speaking are incorporated in GGA, the on-site Coulomb repulsion U was believed to play a very important role mainly to open and keep the proper value of the band gap, and in this way to control the chemical bonding in the system. For example, the metallicity caused by the intrinsic errors of GGA, additionally contributes to the chemical bonding and destroys the JT distortion. Thus, if the material is insulating, this insulating behaviour must be preserved in the electronic structure calculations in order to describe properly the JT effect, and for this purposes the true insulating system cannot be replaced by any auxiliary metallic system in GGA (in the same as the magnetic interactions in insulators cannot modelled by using metallic reference point due to completely different nature of superexchange and double exchange interactions [27]). The canonical example, where GGA completely fails to reproduce any JT instability is KCuF$_3$ [17]. It was concluded that this situation is generic and the on-site Coulomb repulsion U is indispensible in order to describe any JT system. However, less serious problems were encountered later in LaMnO$_3$, where the JT distortion is formally present event at GGA level, but the magnitude of the distortion itself is underestimated [25]. The partial success of GGA for LaMnO$_3$ is related to the fact it becomes insulating if the proper A-type AFM order is taken into account (of course, in order to reproduce the JT distortion in the paramagnetic phase of LaMnO$_3$, it is essential to include the on-site repulsion U and open the band gap [28]). Some critical statements against GGA were relaxed recently. For example, it was shown at once by several groups that GGA is able to reproduce JT distortion in Cs$_2$AgF$_4$ [29-31], which belonged to the same family as KCuF$_3$. Again, the main point is that Cs$_2$AgF$_4$ becomes insulating already at the GGA level.

Our calculations for BiMnO$_3$ also show that distortions of the MnO$_6$ octahedra are very sensitive to correlation effects and the value of the Coulomb repulsion U in the GGA+U method (see Table II). GGA underestimates the experimental JT distortion, mainly due to the additional increase of four short Mn-O distances and decrease of two long one. The best (although not perfect) agreement with experimental data was obtained for U = 2 eV. This value was derived recently for the low-energy Hubbard model, constructed for the 3d-bands of Mn on the basis of first-principles electronic structure calculations [11]. The basic idea is that *3d*-states in manganites, which are admixed into the oxygen 2p-band due to the hybridization effects, participate in the screening of the 3d-states located near the Fermi level. This channel of screening is extremely efficient and substantially reduces the effective value of the Coulomb interaction U in the low-energy region for many transition-metal oxides [32]. In is interesting to note that larger values of U (correspondingly 4 and 8 eV) again decrease the JT distortion, mainly due to the additional (in comparison with U = 2 eV) increase of four short Mn-O distances and expansion of the MnO$_6$ octahedra (Table II). It seems to be reasonable because large U decreases the *3d-2p* hybridization and therefore decreases the bonding effects.

Why BiMnO$_3$ is so distorted and what is the main difference between BiMnO$_3$, LaMnO$_3$,

and BiFeO$_3$, which crystallize correspondingly in the monoclinic, orthorhombic and rhombohedral structure? We speculate that the main difference in the crystal structures is caused by the competition of the JT distortions around Mn-sites and the local off-centrosymmetric distortions around Bi-sites. In the perfect octahedral environment, Mn$^{3+}$ ion has one electron in the double-degenerate $e_g$-shell and this configuration is subjected to the JT instability. Of course, as discussed earlier, it is still an open question how well this instability is reproduced in the electronic structure calculations. However, the effect exists and takes place in three-valent manganites forming the perovskite structure. The chemical role of cations La$^{3+}$, Bi$^{3+}$, etc. is less appreciated. Instead, many properties of perovskites are typically discussed in terms of the size of these three-valent ions, which control the magnitude of the Mn-O-Mn angles and the strength of other distortions in the system. Nevertheless, the situation may be more complicated because both La$^{3+}$ and Bi$^{3+}$ are chemically active and can lead to the off-centosymmetric displacements. Yet, the origin of these displacements is very different for La$^{3+}$ and Bi$^{3+}$.

The formal electronic configuration of Bi$^{3+}$ is 6s$^2$, and the lowest unoccupied states are 6p. Therefore, any off-centosymmetric displacement, which violates the parity, will lower the total energy of the system by mixing the 6s and 6p orbitals. Such a situation does take place in BiMnO$_3$ and BiFeO$_3$ (Fig. 1a) and constitutes the idea of the lone-pair model [8]. This mechanism is essentially local, which acts independently around each Bi-site. It does not say anything about the cooperative character of these displacements, which can either ferro- or antiferroelectric. However, due to its local character, this mechanism is probably very robust.

La$^{3+}$ is the typical d$^0$-ion. In oxides, involving La$^{3+}$ and other d$^0$-ion, the unoccupied d-states (in the present case – the 5d-states of La) form the conduction band, while oxygen 2p-band is fully located in the occupied part of spectrum. Due to the *5d-2p* hybridization, the oxygen *2p-* and La *5d-* bands can be characterized as bonding and antibonding states, respectively. In such a case, the odd-order cooperative displacement, which has finite matrix elements between bonding oxygen *2p-* and antibonding La *5d*-bands, can additionally lower the total energy of the system. Unlike the local displacements around the Bi-sites, this distortion is cooperative and in many cases has ferroelectric character. This mechanism was proposed long time ago by Bersuker as a possible origin of ferroelectric behaviour of $BaTiO_3$ and relates oxide [33], and reviewed in the monograph [34]. However, unlike the local lone pairs in Bi, this mechanism is purely itinerant and not sufficiently strong to compete with the JT effect in magnanites.

Thus, we propose the following picture. Due to the (fully occupied) $d^5$ configuration of $Fe^{3+}$ ions in $BiFeO_3$, it is not a Jahn-Teller system. Therefore, the crystal distortions in $BiFeO_3$ are caused exclusively by local off-centrosymmetric displacements around the Bi-sites and there is no other mechanism, which would compete with this effect. In the local environment of each Bi-site in $BiFeO_3$, there are two *perfect* oxygen triangles, which however are located at different distances from Bi (Fig. 1a). Thus, the inversion symmetry is broken and the system is ferroelectric. $LaMnO_3$, in principle, may have the ferroelectric instability of the itinerant type. However, the effect is probably too weak and easily overpowered by the JT distortion around the Mn-sites. Thus, the main cause of the crystal distortion in $LaMnO_3$ is the regular mismatch of the ionic radii of $La^{3+}$ and $Mn^{3+}$ (the so-called tolerance factor) and the JT effect around the Mn-sites. The uniqueness of $BiMnO_3$ is that in this compound the off-centrosymmetric instability around the Bi-sites readily competes with the JT effect (and the ionic radii mismatch) around the Mn-sites. This is probably the main reason why $BiMnO_3$ forms strongly distorted monoclinic structure.

As was mentioned above, the structural optimization in $BiMnO_3$ decreases the JT distortions (Tables I-III). In the process of optimisation, by starting from the LT structure, three oxygen atoms (O1-O3 in Fig. 1b) move away from the bismuth, while three other oxygen atoms of same classes (O1a-O3a in Fig. 1b) become close to it (Table III). Moreover, the oxygen atoms are shifted in rather inequivalent way: for example, in the more distorted triangle O1a-O3a, two remote atoms O2a and O3a move more strongly than the atom O1a, located closer to Bi. Apparently, these shifts are governed by the Bi-O interactions, which tend to restore the perfect oxygen triangle around each Bi-site (similar to $BiFeO_3$). Since the sites O1a, O2a, and O3a can be transformed to O1, O2, and O3 by the symmetry operations of the C2/c group, some changes are inevitably observed also in the small O1-O2-O3 triangle, where the O1 atom is shifted more strongly away from the central Bi-site in comparison with O2 and O3 (Table III). However, the expansion of the O1-O2-O3 triangle acts against the JT distortions around the Mn-sites and tend to decrease them. Thus, in the process of structure optimization, the balance between the JT distortion around the Mn-sites and off-centrosymmetric displacements around the Bi-sites is shifted towards the latter. We believe that this is the main reason why the JT distortion in the optimized structure is substantially smaller than in the experimental LT structure.

## 4. Defects in BiMnO$_3$

One of the possible reasons why several energetically unfavourable structures of BiMnO$_3$ have been seen in the experiment could be the defects. In order to explore these possibilities, we have calculated the formation energies for LaMnO$_3$, BiFeO$_3$, and BiMnO$_3$. As the starting point for the optimization in the case of BiMnO$_3$, we have used LT structure. As for the defects, we have considered anionic and cationic vacancies as well as the interstitial oxygen. In the latter case, the impurity atom occupies the tetrahedral void (see Figs. 1b and 2), which would correspond to the 3c positions in the perfect cubic environment (having two Bi and four oxygen atoms in the nearest neighbourhood). Contrary to other compounds, the formation energy for interstitial oxygen site in BiMnO$_3$ is rather small (Table IV). Our finding is consistent with experimental report on the lack of the oxygen-deficient samples BiMnO$_{3-x}$ and easy formation of the oxygen-rich ones BiMnO$_{3+x}$ with x varying from 0.04 to 0.25 [10].

In order to explore the concentration and initial state dependence of the oxygen-rich crystal structure of BiMnO$_3$, we have performed the full structural optimization and calculated the formation energies by starting from three experimentally determined structures of BiMnO$_3$ with the large JT distortion, namely A, D and LT, and considered three possible supercells (1×1×1, 2×2×1, 2×2×2), containing 20, 40 and 80 atoms.

There is a number of interesting effects related to the oxygen impurities.

(1) One surprising result is that irrespectively on the starting condition and for all considered concentrations, the defect BiMnO$_{3+x}$ system remains insulating, despite the widespread believe that the impurity oxygen should accept electrons from the occupied Mn *3d* band and turn the system into metallic regime. This apparent contradiction is resolved as follows. It appears that the impurity atom (O$_i$) tends to form a bond with one of the oxygen atoms in the BiMnO$_3$ crystal. For example, in the process of optimization, the O$_i$ atom steadily moves towards the O2 one, so that the O$_i$-O2 distance becomes about 1.41~1.43 Å (where the precise value slightly depends on the concentration and the starting condition). Although this value is larger than O-O distance in the oxygen molecule (1.21 Å), it is substantially smaller than the O-O distance in the MnO$_6$ octahedron and the distance between O$_i$ and other oxygen sites. The formation of the O$_i$-O2 bond is reflected in the behaviour of the density of states (Fig. 3): the density of states for the O2 atom is strongly perturbed and practically repeats the distribution of the impurity O$_i$ states, which indicates at strong interaction between these two atoms. The *6s* states of Bi-atom located near the impurity are shifted down by about 1 eV. Apart from these local effects, other parts of the electronic structure remain practically unchanged. The most important point is that the formal configuration of the newly formed dimer should be O$_2^{2-}$ (the formal configuration of oxygen in pure BiMnO$_3$ is O$^{2-}$ and the impurity atom does not take additional electrons apart from forming the bond). Therefore, the Fermi level should lie in the gap between antibonding molecular levels of the $\pi_g$- and $\sigma_u$-symmetry, which are located in the occupied and unoccupied part of the spectra, respectively. This explains the insulating character of BiMnO$_{3+x}$.

(2) In order to clarify the role played by the oxygen defects in stabilizing large JT distortion, which is seemingly observed in some of the experimental structures [6, 7], we monitor the total energy of the system along different paths connecting the experimental LT structure and optimized crystal structure of BiMnO$_3$, and present it as the function of the Mn1-O2 bondlength, which was most strongly affected by the optimization (Fig. 4). For the defect composition BiMnO$_{3.250}$, we also consider the paths to the "intermediate" configuration, corresponding to the optimized structure with the oxygen defect and monitor. One can clearly see that due to the formation of the pair O2-Oi, the oxygen atom O2 appears to be trapped much closer to the Mn1-site. The energy gain for this process is substantially larger than the one for the oxygen displacements in the pure $BiMnO_3$. This mechanism partially "restores" the JT distortion around the Mn-site, which manifest itself in the existence of two long (>2 Å) and four short (<2 Å) Mn-O bonds (Fig. 2), and can be roughly described by the $Q_3<0$ mode [26]. Nevertheless, the situation around the Mn2-site is different: although there is a strong local distortion, which leads to the additional differentiation of the Mn-O bonds into the short and long ones, it cannot be ascribed to the true type of the JT distortion. Roughly speaking, in this case we have three long (>2 Å) and three short (<2 Å) Mn-O bonds (Fig. 2).

Attempts to place the oxygen impurity in other positions appear to be less energetically favourable: corresponding defect formation energies are at least 50% large in comparison with the configuration shown in Fig. 1b and 2. Moreover, there was no oxygen pair formation and the system was metallic, which is unfavourable for the JT distortion.

(3) We consider the large variety of experimentally observed magnetic structures and argue that some of these structures can be stabilized by the oxygen defects. For these purposes, we start from different experimental structures reported for pure $BiMnO_3$ [5-7] and perform the structural optimization with oxygen defects (see Tab. V). We also consider different concentration of these defects. Surprisingly that depending on the starting configuration we obtain rather different optimized structures, corresponding to different local minima of the total energy $BiMnO_{3+x}$. In this sense, the situation is very different from the pure $BiMnO_3$, where irrespectively on the starting condition we finally arrived at the same optimized structure. Particularly, the oxygen defect formation energies are pretty close for the LT and A structures and decrease with the decrease of x (due to the fact that the defect formation is a local process, which perturbs the crystal structure locally). However, the situation for the D structure is completely different. In this case, for x=0.250 and 0.125 we have obtained negative formation energies, which indicate that the D structure is unstable towards the formation of defects, for the large concentration of the latter. We confirmed that the main reason for such behaviour is related to the local environment of Bi-atoms, which is explained in Fig. 5 (while the Mn-environment in optimized structures is much more similar). Briefly, the oxygen coordination of the Bi-atoms is very different in the experimental A and D structures. Another important difference is bigger unit cell for the D structure, which provides more space for the atomic relaxation. The large difference is observed also in the optimized structures. Particularly, in the optimized D structure with the defect, the Bi-atom has two oxygen triangles in its nearest environment (Fig. 6). Thus, the situation is similar to the one realized in $BiFeO_3$ (Fig. 1a), which is favourable for the Bi-sites. This probably explains the abnormal stability of the oxygen-excessive D-structure.

## 5. Conclusions

Using density-functional calculations, we have argued that highly deformed crystal structure of $BiMnO_3$ is the result of competition of the JT distortion around the Mn-sites and off-centrosymmetric displacements around the Bi-sites, which involve the same oxygen atoms: the former tend to form the JT-distorted ($Q_3<0$ [26]) $MnO_6$ octahedra, while the latter tend form two nonequidistant oxygen triangles around each Bi-site, similar to $BiFeO_3$. The obtained structure appears to be favourable for point defects, especially for the interstitial oxygen impurities, which have the lowest formation energy. The defects lead to the additional distortion and generally destroy the inversion symmetry. They can also lead to the additional differentiation of the short and long Mn-O bonds, and in this sense "mimic" the JT distortion. Nevertheless, the driving force and the symmetry of the distortion are different from the true JT effect. Moreover, the optimization of the oxygen-excessive crystal structure of $BiMnO_3$ appears to be extremely sensitive to the starting conditions. This may explain the fact that it is extremely difficult to synthesise the single crystal of $BiMnO_3$ as well as large variety of experimentally proposed crystal structures of $BiMnO_3$, which may be also affected by the defects.

**Acknowledgements**

We are grateful to A. A. Belik for fruitful discussions. This work is partly supported by Grant-in Aid for Scientific Research (C) No. 20540337 from MEXT, Japan and Russian Federal Agency for Science and Innovations, grant No. 02.740.11.0217.

**Table I.** The values of ratios of the biggest and smallest values of Mn-O distance in manganese octahedrons for all known experimental and optimized atomic structures of BiMnO3, and the values of the total energy differences between studied structures and LT structure (meV/BiMnO$_3$).

|       | A    | D    | LT   | HT   | optimized |
|-------|------|------|------|------|-----------|
| Mn1   | 1.20 | 1.21 | 1.15 | 1.05 | 1.08      |
| Mn2   | 1.21 | 1.30 | 1.17 | 1.10 | 1.07      |
| E-E$_{LT}$ | +327 | +97 | 0 | -38 | -87 |

**Table II.** Values of Mn-O distances (in Å) and ratios of largest and shortest bonds in the optimized crystal structures of BiMnO$_3$, as obtained in the GGA+U calculations with different values of Coulomb repulsion U.

|        | GGA   | 2 eV  | 4eV   | 8eV   |
|--------|-------|-------|-------|-------|
| Mn1-O  | 1.995 | 1.999 | 2.037 | 2.132 |
|        | 2.006 | 2.030 | 2.040 | 2.161 |
|        | 2.093 | 2.155 | 2.090 | 2.144 |
|        | 1.05  | 1.08  | 1.03  | 1.02  |
| Mn2-O  | 1.947 | 1.949 | 1.957 | 2.025 |
|        | 2.002 | 1.977 | 2.019 | 2.053 |
|        | 2.098 | 2.077 | 2.073 | 2.093 |
|        | 1.08  | 1.07  | 1.06  | 1.03  |

**Table III.** Crystallographic data for LT, HT and optimized structure of BiMnO3 (lattice parameters and distances are measured in Å, and angles are measured in degrees).

|        | LT       | HT       | optimized |
|--------|----------|----------|-----------|
| a      | 9.5415   | 9.5866   | 9.6910    |
| b      | 5.6126   | 5.5990   | 5.5697    |
| c      | 9.8630   | 9.7427   | 9.8027    |
| β      | 110.6584 | 108.601  | 110.376   |
| Bi     |          |          |           |
| x      | 0.13638  | 0.13387  | 0.13576   |
| y      | 0.21832  | 0.21603  | 0.22407   |
| z      | 0.12617  | 0.12792  | 0.12358   |
| Mn1    |          |          |           |
| x      | 0        | 0        | 0         |
| y      | 0.21156  | 0.23268  | 0.22703   |
| z      | 0.75     | 0.75     | 0.75      |
| Mn2    |          |          |           |
| x      | 0.25     | 0.25     | 0.25      |
| y      | 0.25     | 0.25     | 0.25      |
| z      | 0.5      | 0.5      | 0.5       |
| O1     |          |          |           |
| x      | 0.09980  | 0.08923  | 0.09241   |
| y      | 0.17233  | 0.18655  | 0.18927   |
| z      | 0.58145  | 0.58863  | 0.58048   |
| O2     |          |          |           |
| x      | 0.14578  | 0.15723  | 0.15079   |
| y      | 0.57143  | 0.55105  | 0.54713   |
| z      | 0.36795  | 0.37443  | 0.37675   |
| O3     |          |          |           |
| x      | 0.35434  | 0.35293  | 0.35179   |
| y      | 0.54843  | 0.54874  | 0.53773   |
| z      | 0.16471  | 0.15863  | 0.16181   |
| Mn1-O2 | 1.906    | 2.011    | 1.999     |
| Mn1-O1 | 1.986    | 2.032    | 2.030     |
| Mn1-O3 | 2.199    | 2.112    | 2.155     |
| Mn2-O3 | 1.924    | 1.913    | 1.949     |
| Mn2-O1 | 1.941    | 2.024    | 1.977     |
| Mn2-O2 | 2.242    | 2.106    | 2.077     |
| Bi-O2  | 2.218    | 2.213    | 2.292     |
| Bi-O1  | 2.239    | 2.304    | 2.352     |
| Bi-O3  | 2.246    | 2.248    | 2.311     |
| Bi-O1a | 2.466    | 2.484    | 2.418     |
| Bi-O3a | 2.710    | 2.751    | 2.651     |
| Bi-O2a | 2.837    | 2.864    | 2.783     |

**Table IV.** Formation energies (in eV) for the vacancies (v) and interstitial (i) defects in $RMO_3$ compounds as obtained in GGA+U with U = 2eV for the 20 atom unit cell.

| Compound | vR | vM | vO | iO |
|---|---|---|---|---|
| $BiMnO_3$ | 3.54 | 4.26 | 2.98 | 0.60 |
| $BiFeO_3$ | 2.83 | 3.75 | 1.67 | 2.86 |
| $LaMnO_3$ | 7.28 | 3.13 | 4.34 | 2.23 |

**Table V.** The formation energies (measured in eV per defect) of oxygen-excessive samples obtained by starting from different experimental $BiMnO_3$ structures and for different concentration of interstitial oxygen defects.

|  | A | D | LT |
|---|---|---|---|
| $BiMnO_{3.250}$ | 0.57 | -0.41 | 0.60 |
| $BiMnO_{3.125}$ | 0.43 | -0.66 | 0.34 |
| $BiMnO_{3.065}$ | 0.29 | 0.55 | 0.26 |

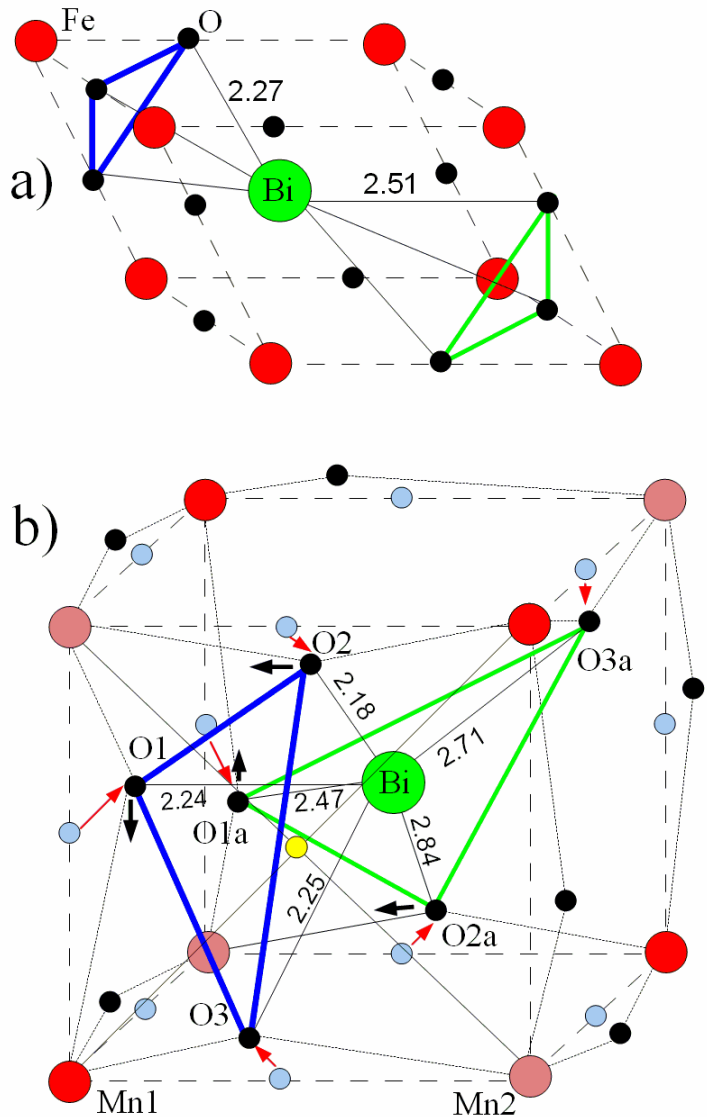

**Figure 1** Schematic view on the crystal structure of BiFeO$_3$ (a) and BiMnO$_3$ (b). Three nearest to bismuth oxygen atoms are denoted by the blue triangle, next nearest three oxygen atoms are denoted by the green triangle. On the panel (b), the positions of oxygen atoms in the ideal perovskite structure are shown by the light blue circles, the shift from the ideal positions to the ones in realistic low-temperature (LT) structure are shown by the red arrows and further shift corresponding to transition to the high temperature (HT) phase is shown by black arrows. The interstitial oxygen atom is denoted by yellow circle. All values are measured in Angstroms. The numbering of Mn- and O-atoms is the same as in Tables I-III.

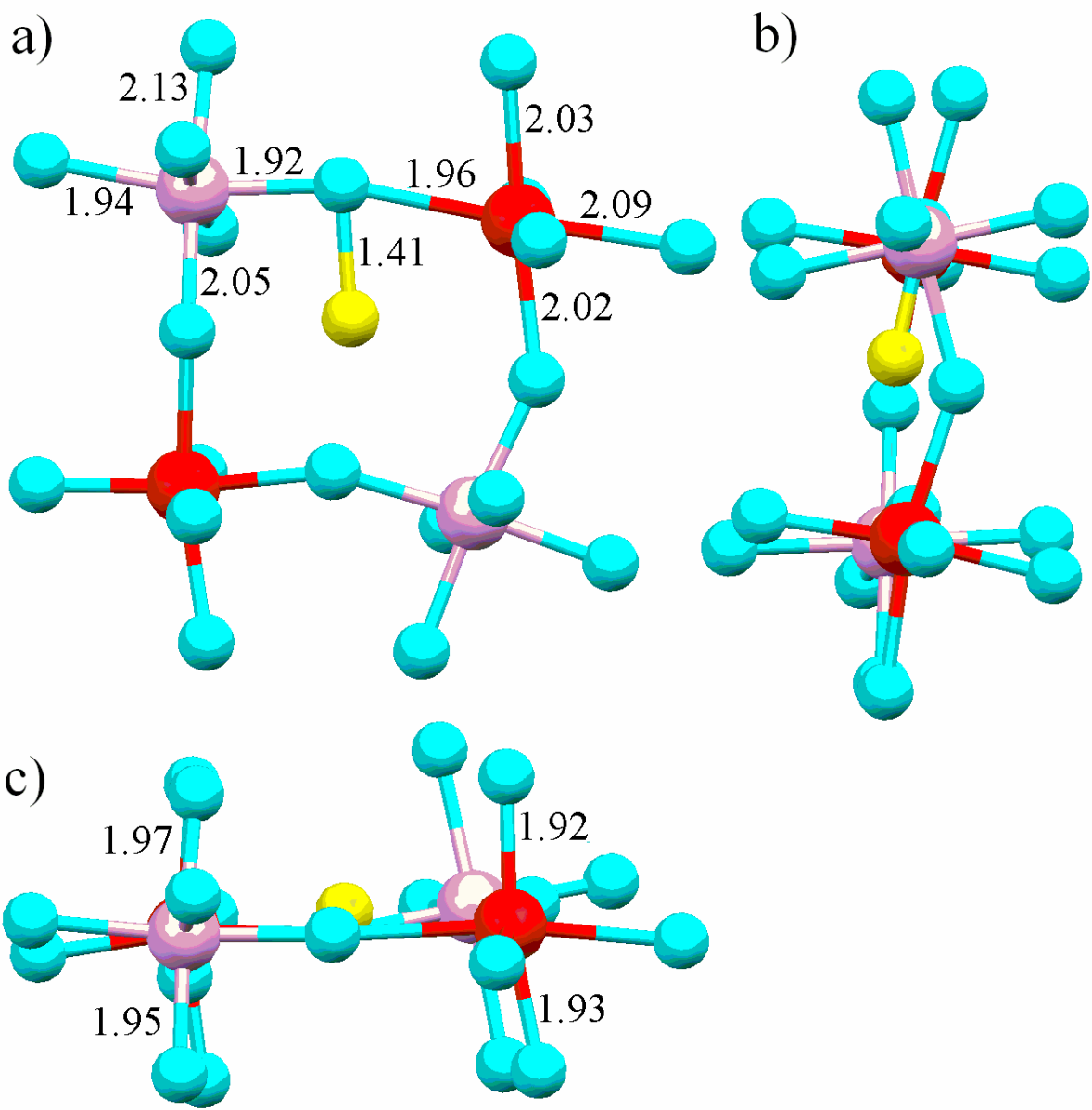

**Figure 2.** Front (a), side (b) and top (c) view of the optimized atomic structure of BiMnO$_{3.250}$ near interstitial oxygen impurity (yellow sphere). Mn1 and Mn2 atoms are indicated by the violet and red spheres, respectively, and oxygen atoms indicated by the blue spheres. All values are measured in Angstroms.

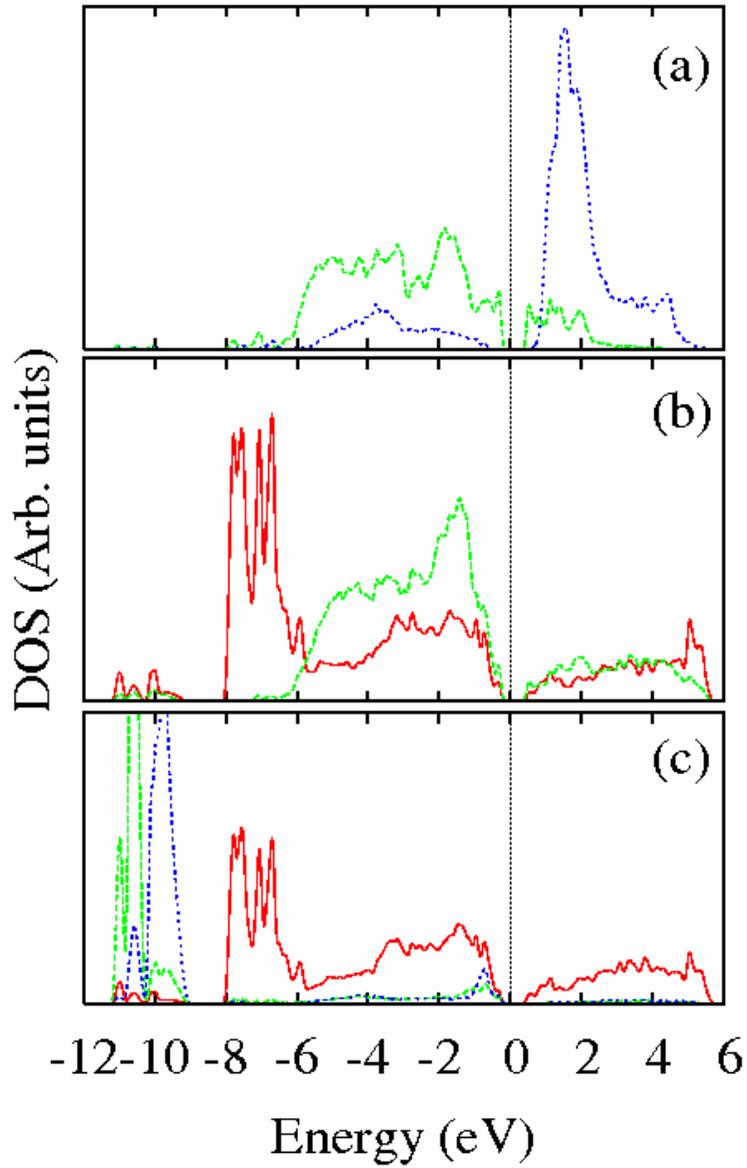

**Figure 3.** Densities of states for BiMnO$_{3.250}$: (a) Mn *3d* states corresponding to the spin-up (dashed green line) and spin-down (dotted blue line); (b) O *2p* states of the O2 atom located near the interstitial oxygen site (solid red line) and of the O2a atom belonging to the same class; (c) O *2p* states of the interstitial oxygen site (solid red line), and Bi *6s* states of the bismuth atoms located near (dashed green line) and far (dotted blue line) from the interstitial oxygen site.

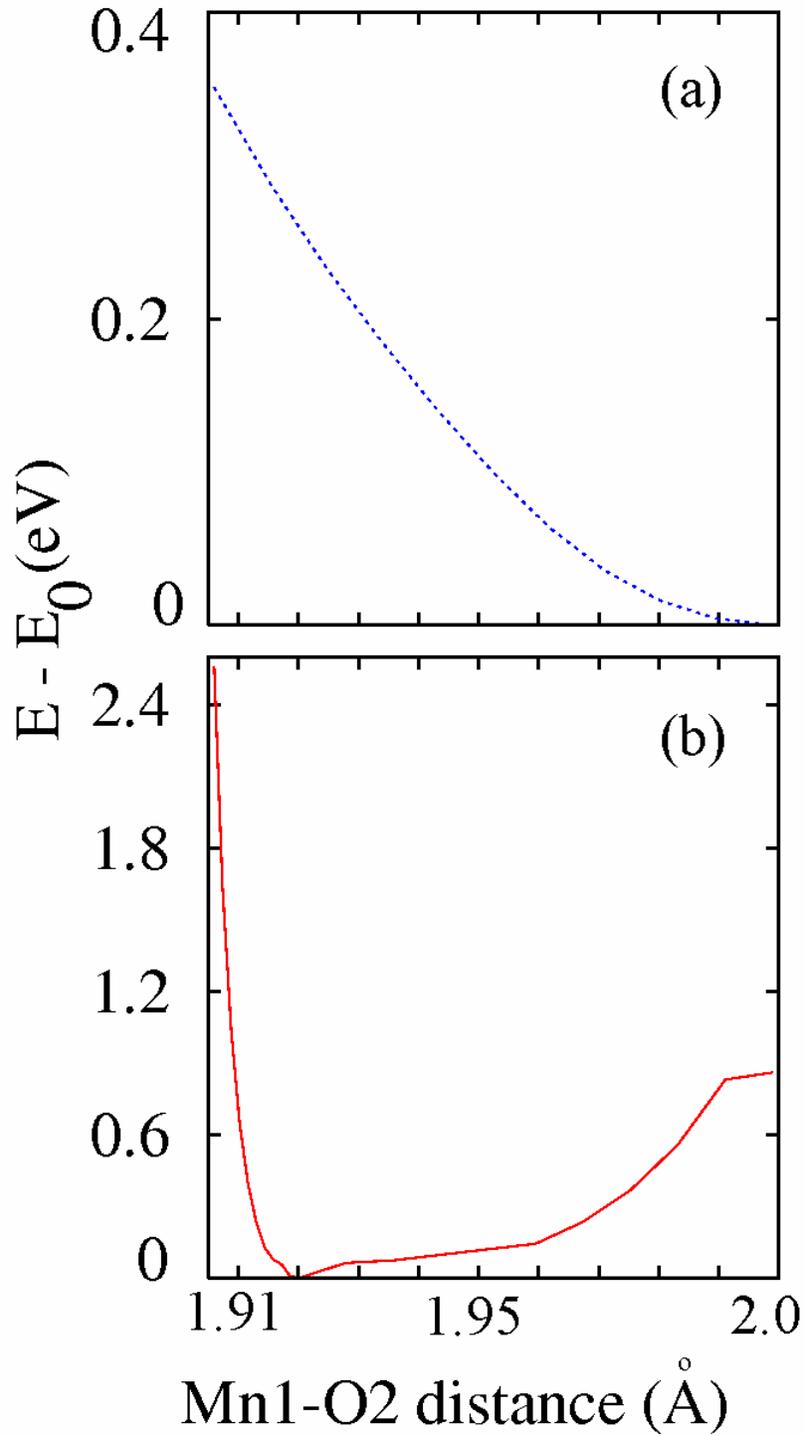

**Figure 4.** (a) Change of the total energy of pure BiMnO$_3$ along the continuous path connecting the experimental LT (left side) and optimized (right side) crystal structure presented as the function of the Mn1-O2 distance (see Fig. 1b for the notations). (b) similar calculations for BiMnO$_{3.250}$ performed via intermediate point, corresponding to the optimized crystal structure with the oxygen defect, which is also used as the new reference point for the total energy.

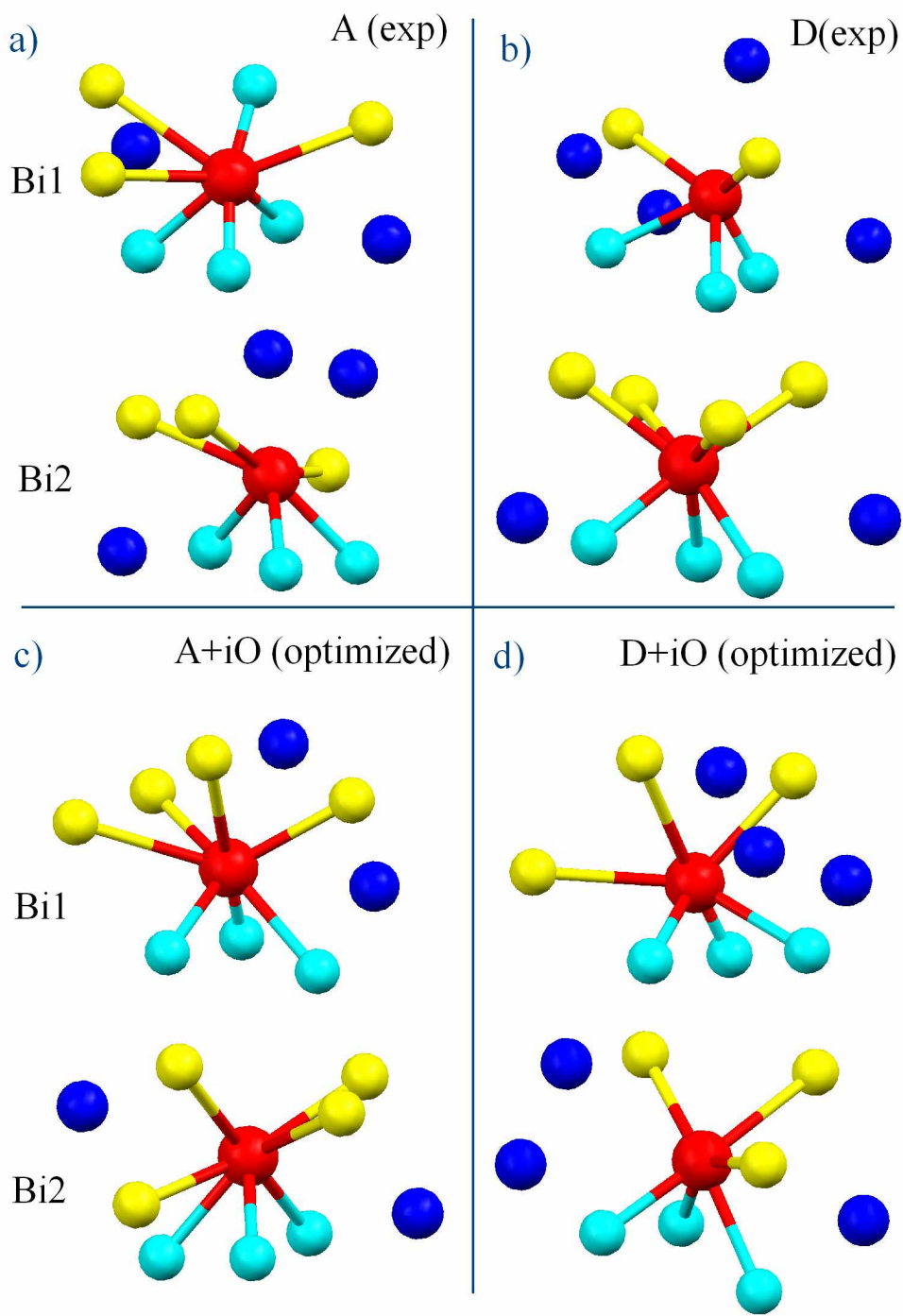

**Figure 5.** Local oxygen environment around inequivalent bismuth atoms (shown by red spheres) for two experimentally observed BiMnO$_3$ structures: (a, Ref. [Atou]) and (b, Ref. [dS]); and for two optimized BiMnO$_{3.25}$ structures: (c) and (d), obtained by starting from the initial atomic configurations (a) and (b), respectively. The oxygen atoms located in the first, second, and third coordination spheres are indicated by the light blue, yellow, and dark blue spheres, respectively.